\documentclass[11pt]{article}
\usepackage[utf8]{inputenc} % Required for inputting international characters
\usepackage[T1]{fontenc} % Output font encoding for international characters

\usepackage{graphicx}
\usepackage{amsmath}
\usepackage{amssymb}
\usepackage{slashed}
\usepackage{mathrsfs}
\usepackage{amsfonts}
\usepackage{braket}
\usepackage{hyperref}
\usepackage{cancel}
\usepackage{cleveref}
\usepackage{dsfont}
\usepackage{wasysym}
\usepackage{subcaption}
\usepackage{tikz-feynman}
\usetikzlibrary{decorations.pathmorphing}
\newcommand{\dd}{\mathop{\mathrm{d}\!}{}}

\newcommand{\deriv}[2]{\dfrac{\dd #1}{\dd #2}}

\newcommand{\HH}{\mathcal{H}}
\newcommand{\LL}{\mathcal{L}}
\newcommand{\OO}{\mathcal{O}}

\newcommand{\bone}{\mathds{1}}
\newcommand{\eqdef}{\stackrel{\text{def}}{=}}

\newcommand{\T}[1]{\mathcal{T}[#1]}

\tikzset{snake it/.style={decorate, decoration=snake}}

\usepackage[backend=bibtex,style=numeric-comp,sorting=none,natbib=true]{biblatex} % Use the bibtex backend with the authoryear citation style (which resembles APA)

\usepackage{bm}
\let\vec\bm

\def\KL{K\"allen-Lehmann\,}
\def\Ep{E_{\vec p}}

\makeatletter

\@addtoreset{equation}{section}

\makeatother
\title{A Note on the Existence of Equal Time Correlators}
\author{Bruno Bucciotti}

\begin{document}

\setcounter{page}{0}
\thispagestyle{empty}

\parskip 3pt

\font\mini=cmr10 at 2pt

\begin{titlepage}
	~\vspace{1cm}
	\begin{center}
		
		\vspace*{-.6cm}
		
		\begin{center}
			
			\vspace*{1.1cm}
			
			{\centering \Large\textbf{A Note on the Existence of Equal Time Correlators}}
			
		\end{center}
		
		\vspace{0.8cm}
		{\bf Bruno Bucciotti$^{a,b}$}

		\vspace{1.cm}
		
		${}^a\!\!$
		{\em  Scuola Normale Superiore, Piazza dei cavalieri 7, 56126 Pisa, Italy}
		
		\vspace{.3cm}
		
		${}^b\!\!$
		{\em INFN, Sezione di Pisa, Largo B. Pontecorvo, 3, 56127 Pisa, Italy}
		
	\end{center}
	
\begin{abstract}
    The Schroedinger picture, which underpins the Wavefunction of the Universe framework to compute Cosmological Correlators, is known to be generically problematic in QFT because of the required infinite localization of the fields in time. We study under which conditions momentum space equal time correlators of scalar fields are finite in flat space. We identify cases where they can be divergent even after renormalizing the theory, while also providing sufficient conditions for their existence. Concrete examples are discussed, covering the well known $\lambda\phi^4$ model, composite operators and effective field theories.
\end{abstract}
\end{titlepage}

\tableofcontents

\section{Introduction}
It is believed that at its very beginning, the universe underwent a phase of rapid expansion named inflation. Describing the physics of inflation demands that we understand quantum field theory on a de~Sitter background, a task to which significant effort has been devoted since the seminal work \cite{Mukhanov:1981xt}. In particular, cosmological correlators have recently attracted a lot of attention (\cite{Flauger:2022hie,Arkani-Hamed:2017fdk,Bzowski:2020kfw,Arkani-Hamed:2015bza,Arkani-Hamed:2018kmz,Meltzer:2021zin,Sleight:2019mgd,Goodhew:2020hob,Benincasa:2020aoj,Jazayeri:2021fvk,Lee:2023kno,Farrow:2018yni,Armstrong:2020woi,Albayrak:2020fyp,Jain:2021qcl,Herderschee:2022ntr,Cheung:2022pdk,Armstrong:2023phb,Mei:2023jkb,Gomez:2021qfd,Sleight:2019hfp,Bzowski:2012ih,Bzowski:2013sza,Bzowski:2017poo,Bzowski:2018fql,Heckelbacher:2020nue,Heckelbacher:2022hbq,Cacciatori:2024zrv}). The aim of the program is to compute $n$\nobreakdash-point correlation functions $\braket{\phi_1\dots\phi_n}$ in a (asymptotically\nobreakdash-)~de~Sitter background \cite{Maldacena:2002vr,Weinberg:2005vy}. An especially successful tool for this purpose is the Wavefunction of the Universe (WFU) \cite{Hartle:1983ai,Maldacena:2002vr,Arkani-Hamed:2015bza,Arkani-Hamed:2018kmz}, see also \cite{Cespedes:2020xqq,Bonifacio:2021azc,Bonifacio:2022vwa,Meltzer:2021zin,Goodhew:2020hob,Goodhew:2021oqg,Hillman:2021bnk,Jazayeri:2021fvk,Melville:2021lst,Lee:2023kno}. An important feature of this approach is that, while it circumvents the more cumbersome computations of the in\nobreakdash-in Schwinger-Keldysh formalism \cite{Chen:2017ryl,Heckelbacher:2022hbq,Chowdhury:2023arc,Beneke:2023wmt,Donath:2024utn}, it most naturally gives us access only to \emph{equal} time correlators. Additionally, even the in\nobreakdash-in formalism is often specialized to the computation of equal time correlators.
The purpose of this work is to critically examine the existence of these equal time correlators in a quantum field theory.
Among the recent developments in the field of cosmological correlators, we highlight that the analog of the Cutkosky cutting rules and the optical theorem were recently formulated for a de~Sitter (actually FLRW) background \cite{Meltzer:2021zin,Goodhew:2020hob,Goodhew:2021oqg,Baumann:2021fxj,Jazayeri:2021fvk,Melville:2021lst,Lee:2023kno,Goodhew:2021oqg} in the language of the WFU, qualifying as all loop orders results. On the fully non-perturbative side, we bring the reader's attention to a formulation of the \KL representation for dS \cite{Bros:1994dn,Cacciatori:2024zrv,Loparco:2023rug}, of the Reeh-Schlieder theorem \cite{Bros:1995js} and to a discussion of the spectral condition \cite{Moschella:2024kvk}.

In this work we examine the main objects of interest that the WFU approach aims to compute, namely the equal time correlators, where currently UV divergences are often not renormalized but an explicit cutoff dependence is kept in the final result (see however \cite{Senatore:2009cf,Heckelbacher:2022hbq,Chowdhury:2023arc,Beneke:2023wmt,Negro:2024bbf,Negro:2024iwy} for examples of works that do take renormalization into account). Following textbook arguments, we will first demonstrate on general grounds that computing correlators at exactly equal time can lead to divergences \emph{even for renormalized fields} at loop level, making the removal of the UV regulator problematic in general. Since we will only be discussing UV divergences, we find it acceptable to move the discussion from de~Sitter to flat space, where both theory and computations are much more developed, on the grounds that the UV behavior of the correlators should be similar.

In \cref{sec:higher_point_functions_theorem} instead we present a positive result, which guarantees the existence of equal time $n$\nobreakdash-point functions under assumptions that can be checked on the sole $2$\nobreakdash-point function. A possible way to sidestep the divergences is presented in \cref{sec:solution}.
We will then present explicit computations in flat space (only touching on de~Sitter) to show how these divergences arise in perturbative computations (\cref{sec:examples}), discussing composite operators and non-renormalizable theories as well.

While the end goal of the program is to extract \emph{cosmological correlators}, which are the analog of scattering amplitudes for de~Sitter, the limit $\eta\rightarrow0$ of the correlators is sometimes plagued by IR divergences. While our understanding of these divergences has significantly improved (see \cite{Seery:2010kh,Bros:2010rku,Burgess:2010dd,Giddings:2010nc,Senatore:2012nq,Akhmedov:2017ooy,Cespedes:2023aal,Gorbenko:2019rza,Mirbabayi:2019qtx,Cohen:2020php,Wang:2022eop} for a limited selection of papers), we will simplify matters by restricting the discussion to finite $t$ or $\eta$. For simplicity, we will also restrict to massive scalar fields, although there is no conceptual difficulty in increasing spin. The mass qualifies as an IR regulator, but will otherwise play no role.

\paragraph{Notation}: We denote bare fields with a subscript $0$, while renormalized fields are denoted by a subscript $R$, or no subscript at all. For equal time (E.T.) correlators, we indicate the time $t$ as $\braket{\dots}_t$. By $\braket{\dots}'$ we mean the correlator without the overall $3-$ or $4-$momentum conserving delta function, depending on context. Our metric convention is $+---$. Except for \cref{sec:general_argument}, we will leave the time ordering symbol implicit in the correlators.

\section{General arguments}
\label{sec:general_argument}
In perturbation theory, an interacting QFT is built starting from bare, UV regulated fields $\phi_0,\,\pi_0$ satisfying equal time canonical commutation relations (CCR), computing correlation functions, and finally rescaling bare fields and couplings to make correlators finite as the regulator is removed. More precisely, we should require \emph{smeared} correlators to be finite~\cite{Streater:1989vi}
\begin{equation}
\label{eq:smeared_correlator}
	\braket{\phi_1(f_1)\dots\phi_n(f_n))} \eqdef \int \dd^4 x_1\dots\dd^4 x_n\, f_1(x_1) \dots f_n(x_n) \braket{\phi_1(x_1)\dots\phi_n(x_n)}
\end{equation}

where the smearing is also in each of the (independent) times the correlator depends on. Except for free field theories, this time smearing is not optional. To see this, we recall an argument given in section 2.4 of \cite{Strocchi:2013awa}. From the CCR we have
\begin{equation}
	\label{eq1:bare_commutator}
	[\phi_0(\vec x,t), \pi_0(\vec y,t)] = i \delta^3(\vec x-\vec y)
\end{equation}
The bare and renormalized fields are linked by a rescaling coefficient as $\phi_0 = Z^{1/2}\phi_R$, giving
\begin{equation}
    [\phi_R(\vec x,t), \pi_R(\vec y,t)] = i Z^{-1}\delta^3(\vec x-\vec y)
\end{equation}
In the simplest case of no derivative interactions, this implies
\begin{equation}
	\label{eq1:renorm_commutator}
	[\phi_R(\vec x,t),\dot\phi_R(\vec y,t)] = i Z^{-1} \delta^3(\vec x-\vec y)
\end{equation}
where $Z^{-1}$ diverges as we remove the regulator, except for free theories where $Z=1$. Thus either $\braket{\phi_R(\vec x)\dot\phi_R(\vec y)}_t$ or $\braket{\dot\phi_R(\vec y)\phi_R(\vec x)}_t$ (in fact both) must be divergent, even after a space smearing in $\dd^3x\dd^3y$ which regulates the $\delta$ function. This divergence only appears if the test functions have intersecting support, suggesting that the problem lies at coincident points.

We will see that the problem appears because of the vanishing time difference between the two fields, so averaging in the overall time $t$ does not help. In fact, time translations make the correlator independent of the overall $t$ in flat space.

This result could seem puzzling because we are discussing correlation functions of renormalized fields, after all divergences have supposedly been subtracted. We will review below, explicitly in section~\ref{sec:phi4_flat}, how this divergence originates.

In the rest of this section we will first describe the condition for the existence of the 2\nobreakdash-point function at equal time (E.T.), guiding the reader's intuition by physically interpreting the result in a CFT. Then we study what happens if we introduce a time smearing which is then sent to a delta function. We conclude by proving a powerful result stating that the finiteness of the E.T. 2\nobreakdash-point function implies the finiteness of all E.T. n-point functions.

\subsection{Discussion of the 2-point function}
\label{sec:KL-general}
We can discuss the finiteness of the 2\nobreakdash-point function at equal times by invoking the \KL representation \cite{Weinberg:1995mt,Strocchi:2013awa} (for now we don't specify whether $\phi(x),\rho(\mu^2)$ are bare or renormalized). We discuss the time ordered correlator, but the result easily translates to Wightman functions.
\begin{equation}
\label{eq:Tordered_2-pt_KL}
	\braket{\T{\phi(x)\phi(y)}} = \int_0^\infty \dd\mu^2\, \rho(\mu^2) D(x-y;\mu^2) = \int\dd\mu^2\frac{\dd^4p}{(2\pi)^4}\,\rho(\mu^2)\, \frac{i}{p^2-\mu^2+i\varepsilon} e^{ip(x-y)}
\end{equation}
where $D$ is the Feynman propagator, $\rho$ is the spectral density and $\mu$ is the mass of the intermediate states. We stress that this equation should be understood in the distributional sense, so first of all one would integrate in spacetime against some test function as in \cref{eq:smeared_correlator}, then one integrates in the momentum, and only then in $\mu^2$.

We will now determine under which hypothesis the idea of having all fields at equal times (E.T.) can work, while keeping the space smearing against a test function $f(\vec x)$. In agreement with causality, we integrate in $E$ by closing the contour in the upper/lower half plane, depending on the sign of $\Delta t$. We get
\begin{align}
\label{eq2:flat_space_2correlator}
\begin{split}
	\braket{\T{\phi(f,t_1)\phi(f,t_2)}} = \int\dd\mu^2\rho(\mu^2)\int\frac{\dd^4 p}{(2\pi)^4}\, \frac{i}{E^2-\vec p^2-\mu^2+i\varepsilon} |\tilde f(\vec p)|^2 e^{iE\Delta t} =\\
	= \frac{1}{2} \int\dd\mu^2\rho(\mu^2)\int\frac{\dd^3 p}{(2\pi)^3} \frac{|\tilde f(\vec p)|^2}{\Ep} e^{-i\Ep |\Delta t|},\quad \Ep = \sqrt{\vec p^2+\mu^2},\quad \Delta t = t_1-t_2
\end{split}
\end{align}
If we are interested in the correlator at some given $3-$momentum $\vec p$ (amputating the momentum conserving $\delta^3$ function), it is sufficient to remove $\int\frac{\dd^3 p}{(2\pi)^3} |\tilde f(\vec p)|^2$. The Wightman function would instead be
\begin{equation}
\label{eq:Wightman_2-pt_KL}
    \braket{\phi(f,t_1)\phi(f,t_2)} = \frac{1}{2} \int\dd\mu^2\,\rho(\mu^2)\int\frac{\dd^3 p}{(2\pi)^3} \frac{|\tilde f(\vec p)|^2}{\Ep} e^{-i\Ep \Delta t}
\end{equation}
For E.T. we then get a finite 2\nobreakdash-point function if and only if
\begin{equation}
    \braket{\phi(f,t)\phi(f,t)} = \frac{1}{2} \int\dd\mu^2\,\rho(\mu^2)\int\frac{\dd^3 p}{(2\pi)^3} \frac{|\tilde f(\vec p)|^2}{\Ep} <\infty
\end{equation}
Since the integrand is always positive, we need to ensure convergence in $\mu^2$. Recalling $E_{\vec p}\sim \mu$ as $\mu\rightarrow\infty$, we derive our first result that the 2\nobreakdash-point function at E.T. is finite if
\begin{equation}
    \label{eq2:existence_of_two_point_function}
    \int\dd\mu^2\frac{\rho(\mu^2)}{\mu}<\infty
\end{equation}
Up to now our discussion applies for any field. We now specialize the discussion to the \textquotedblleft elementary\textquotedblright{} fields that appear in the Lagrangian. We will show that the correlator between a renormalized field and its conjugate momentum is divergent at equal times, in agreement with the general argument above.

Going back to~\cref{eq2:flat_space_2correlator}, working at fixed $3-$momentum for simplicity, observe that if we take the time derivative of one of the fields, convergence worsens because of the extra factor $\Ep$ that is pulled down. This is very explicit if we compute
\begin{equation}
    \label{eq2:phi_phi_dot_unequal_t}
	\braket{\T{\phi(\vec p,t_1)\dot\phi(\vec p,t_2)}}' = \frac{i}{2} sgn(t_1-t_2) \int\dd\mu^2\,\rho(\mu^2)\, e^{-i\Ep |\Delta t|}
\end{equation}
which could be problematic when $t_1\rightarrow t_2$\footnote{This is so regardless of the sign of $t_1-t_2$. Actually, the prefactor $sgn(t_1-t_2)$ comes from time ordering, and would be absent for the Wightman function, thus posing no conceptual problem.}. To see that $\int\dd\mu^2\rho(\mu^2)$ is divergent, we will relate it to the well known commutator of bare fields.

Given that~\cref{eq2:phi_phi_dot_unequal_t} is the time-ordered correlator, we can take the difference between the limits $t_1\rightarrow t_2^+$ and $t_1\rightarrow t_2^-$, which (setting $t_2=t$) effectively computes the commutator $[\phi(\vec p,t),\dot\phi(\vec p, t)]$.
This computation can be specialized to bare or renormalized fields, giving
\begin{equation}
	\braket{[\phi_R,\dot\phi_R]}_t' = i\int\dd\mu^2\,\rho_R(\mu^2),\qquad
	\braket{[\phi_0,\dot\phi_0]}_t' = i\int\dd\mu^2\,\rho_0(\mu^2),\quad \rho_0 \eqdef Z \rho_R
\end{equation}
The momentum field $\pi$ conjugate to $\phi$ is given by $\dot\phi$, up to corrections that vanish at weak coupling. The bare fields are canonically quantized (this is at the basis of the Feynman perturbative expansion),
\begin{equation}
    \braket{[\phi_0(\vec p,t),\pi_0(\vec p,t)]}_t' = i
\end{equation}
Thus from $\braket{[\phi_0,\dot\phi_0]}_t' = i\int\dd\mu^2\,\rho_0(\mu^2)$ we have
\begin{equation}
	\label{eq2:rho0_integral_1}
    \int\dd\mu^2\rho_0(\mu^2)=1
\end{equation}
and we deduce our second result
\begin{equation}
	\label{eq2:rho_integral_divergence}
	\int\dd\mu^2\,\rho_R(\mu^2)=Z^{-1}
\end{equation}
which is divergent as we remove the cutoff. This result indicates that not only the E.T. commutator of $\phi$ and its conjugate momentum is divergent at E.T., but \cref{eq2:phi_phi_dot_unequal_t} at E.T. as well (regardless of $sgn(t_1-t_2)$). This implies that both $\braket{\phi_R\dot\phi_R}_t'$ and $\braket{\dot\phi_R\phi_R}_t'$ are divergent at E.T. .

\subsection{Interpretation in a CFT}
\label{sec:interpretation}
In this subsection we will try to give an interpretation of the condition in \cref{eq2:existence_of_two_point_function} that determines if the 2\nobreakdash-point function exists at equal times. We will do so by studying an example, the 2\nobreakdash-point correlator for a (scalar) field in a conformal theory, with dimension $\Delta$. The reason why this example is simple is that, from scaling arguments, $\rho(\mu^2)~\propto~(\mu^2)^{\Delta-2}$, so the integral in \cref{eq2:existence_of_two_point_function} is convergent if and only if
\begin{equation}
\label{eq:existence_2pt_cft_condition}
    \Delta<\frac{3}{2}
\end{equation}
Incidently, the scaling arguments apply even for spinning fields, which however require a separate discussion for the edge case $\Delta=\frac{3}{2}$ due to the non\nobreakdash-trivial tensor structures that appear.

Recall that in Euclidean signature the 2\nobreakdash-point function of a scalar field of dimension $\Delta$ is
\begin{equation}
\label{eq:cft_euclidean}
    \braket{\phi(\tau_1,\vec x_1)\phi(\tau_2,\vec x_2)} = \frac{1}{[(\tau_1-\tau_2)^2+(\vec x_1-\vec x_2)^2]^{\Delta}}
\end{equation}
Notice how the only singularity is localized at coincident points. Instead in Lorentzian signature, we get the Wightman 2\nobreakdash-point function
\begin{equation}
\label{eq:cft_lorentzian}
    \braket{\phi(t_1,\vec x_1)\phi(t_2,\vec x_2)} = \frac{1}{[-(t_1-t_2-i\varepsilon)^2+(\vec x_1-\vec x_2)^2]^{\Delta}}
\end{equation}
where singularities, now spread out on the whole light-cone, are regulated by $i\varepsilon$. The appearence of $i\varepsilon$ is most easily understood by looking at \cref{eq:Wightman_2-pt_KL}: we can either view it in the distributional sense, but we can also send $\Delta t\rightarrow\Delta t-i\varepsilon$ obtaining an analytic function in the complex time \cite{Streater:1989vi}.

We now come to what happens when we consider equal times. The 2\nobreakdash-point function becomes
\begin{equation}
\label{eq:cft_ET}
    \braket{\phi(t,\vec x_1)\phi(t,\vec x_2)} = \frac{1}{[(\vec x_1-\vec x_2)^2]^{\Delta}}
\end{equation}
which looks fine, except at coincident points. However, to be more precise, \cref{eq:cft_ET} is not a distribution in general. To see this, it is sufficient to study the existence of the integral of \cref{eq:cft_ET} against a test function in space\footnote{See for example \cite{Gillioz:2022yze}, pag. 25, 28, for a brief discussion of test functions.}. The integral we study is
\begin{equation}
    \int \frac{1}{[(\vec x)^2]^{\Delta}} \dd^3x
\end{equation}
where we assumed the test function to have non-zero value close to $\vec x=0$. The result is indeed finite if and only if $\Delta<3/2$, in agreement with \cref{eq:existence_2pt_cft_condition}. Since the problem we are highlighting has to do with the UV modes of the theory, and given that many quantum field theories are expected to flow to a conformal fixed point in the UV, the computation above actually applies quite generally.

While \cref{eq2:existence_of_two_point_function,eq2:rho_integral_divergence} were indicating a problem with large momenta, this discussion highlights that the issue has to do with coincident points, thus offering a complementary perspective. As we will explore in more detail later, one implication of the fact that \cref{eq:cft_ET} is not a well-defined distribution is that this correlator does not admit a Fourier transform to momentum space.
\medskip

Extensions to spinning fields are also possible. For example, for spin $\frac{1}{2}$ fermions of dimension $\Delta$, the bound in \cref{eq:existence_2pt_cft_condition} becomes $\Delta\le\frac{3}{2}$. In addition, unitarity bounds for spinning particles~\cite{Mack:1975je} enforce $\Delta\ge\frac{3}{2}$, pinning $\Delta$ to $\frac{3}{2}$ and thus reducing the class of fermion fields that admit a sharp time restriction down to free fermions only. This result was also derived by Wightman~\cite{Wightman:1967kq} by different means.

\subsection{Time smearing}
\label{sec:time_smearing}
We know from the general theory of Wightman QFT that renormalized correlators give finite numbers when we smear all coordinates using test functions regular in all space-time variables, as in \cref{eq:smeared_correlator}. We can then ask what goes wrong as we send the time regulator to a sharp $\delta$ function. What we will see is that the time smearing cuts off dangerous UV modes whenever it is present, and taking it to a $\delta$ function removes this crucial suppression.

To make matters precise, we will commit to a particularly easy Gaussian smearing, but these results extend as long as the test functions have support over some characteristic time/length scale. We choose the normalized
\begin{equation}
    \alpha(t) = \frac{\Omega}{\sqrt{\pi}}e^{-\Omega^2t^2},\qquad
    f(\vec x) = \frac{K^2}{\pi^{3/2}}e^{-K^2|\vec x|^2}
\end{equation}
so that \cref{eq:Wightman_2-pt_KL} becomes
\begin{equation}
    \braket{\phi(f,\alpha)\phi(f,\alpha)} = \int_0^\infty\dd\mu^2\rho(\mu^2)\int\frac{\dd^3p}{(2\pi)^3}\frac{1}{2E_{\vec p}}e^{-\frac{E_{\vec p}^2}{2\Omega^2}}e^{-\frac{\vec p^2}{2K^2}}
\end{equation}
We again stress the order of the integrals in \cref{eq:Tordered_2-pt_KL}: first in $x,\,y$ against test functions, then in momentum, then in $\mu^2$. The integral in $\vec p$ gives, up to irrelevant scaleless prefactors,
\begin{equation}
    \int_0^\infty\dd\mu^2\rho(\mu^2) \frac{K^2\Omega^2}{K^2+\Omega^2}e^{-\frac{\mu^2}{2\Omega^2}}U\left(\frac{1}{2},0,\mu^2\frac{K^2+\Omega^2}{2K^2\Omega^2}\right)
\end{equation}
where $U$ is the confluent hypergeometric funtion, which asymptotes to $U(\frac{1}{2},0,x)\sim\frac{1}{\sqrt{x}}$ as $x\rightarrow\infty$. The well known sub-exponential growth of $\rho(\mu^2)$, together with the exponential suppression from $e^{-\frac{\mu^2}{2\Omega^2}}$ now ensures the convergence of the integral for any $\Omega>0$.

We can easily study what happens in the limit in which $\alpha(t)\rightarrow\delta(t)$, i.e. $\Omega\rightarrow\infty$: because the integrand is positive, no non-trivial cancellations can arise. For this reason, we can bring the limit inside the integral and we get
\begin{equation}
    \int_0^\infty\dd\mu^2\rho(\mu^2) K^2 U\left(\frac{1}{2},0,\frac{\mu^2}{2K^2}\right)
\end{equation}
which, recalling the asymptotic behavior of $U$, is convergent if and only if
\begin{equation}
    \int^\infty\dd\mu^2\frac{\rho(\mu^2)}{\mu}<\infty
\end{equation}
in agreement with \cref{eq2:existence_of_two_point_function}.

\subsection{(Renormalized) Higher point functions}
\label{sec:higher_point_functions_theorem}
Finiteness is relatively easy to establish for the 2\nobreakdash-point function, more than for any of the higher point functions, because only then do we have the \KL representation. Luckily, finiteness of the 2\nobreakdash-point function turns out to be enough to guarantee finiteness of any n-point function. In the following discussion we will only work with renormalized fields, for they are the only ones that give finite correlation functions when appropriately smeared.

To show this, we begin by establishing that $\phi(f,t)=\int\phi(\vec x,t)f(\vec x)\dd^3 x$ (i.e. the field restricted to sharp time but smeared in space) is a well defined operator on a dense subspace of $\HH$ when the 2\nobreakdash-point function at E.T. is finite. Then, to conclude the argument, we simply observe that the $n$\nobreakdash-point function of $\phi$ at equal time is just the (repeated) application of a well defined operator, $\phi(f,t)$, on a state in $\HH$, so no divergence can arise. For simplicity we neglect possible domain issues that could arise for such non-compact operators, effectively blurring the distinction between a state and an arbitrarily close approximation of it.

\paragraph{Argument}
We begin by proving that if the E.T 2\nobreakdash-point function exists, then $\phi(f,t)$ is a well defined operator.
If $\braket{\phi(f,t)\phi(g,t)}$ is finite for any space test functions $f,g$, then $\phi(f,t)\ket{0}$ is a well defined state because it has finite norm (specializing to $g=f$). Assume that $f$ has support in some finite space region $R_1$. We now want to determine the matrix elements of $\phi(f,t)$ itself, and we will do so by considering the braket
\begin{equation}
    \braket{\psi|\phi(f,t)|\phi}
\end{equation}
By the Reeh-Schlieder theorem\footnote{See \cite{Streater:1989vi,Strocchi:2013awa} for a discussion of the Reeh-Schlieder theorem and some of its applications.}, we can approximate (arbitrarily well) $\ket{\phi}$ by acting on the vacuum with fields localized in any spacetime region (even compact!) we desire. Let
\begin{equation}
    \ket{\phi} = \Phi(R_2)\ket{0}
\end{equation}
where $\Phi(R_2)$ are fields smeared in the spacetime region $R_2$, which we choose to be space-like with respect to $R_1$ localized at time $t$. Then
\begin{equation}
    \braket{\psi|\phi(f,t)|\phi} = \braket{\psi|\phi(f,t)\Phi(R_2)|0}
\end{equation}
but from microcausality of the fields (commutation at space-like separation) we have
\begin{equation}
    \braket{\psi|\phi(f,t)\Phi(R_2)|0} = \braket{\psi|\Phi(R_2)\phi(f,t)|0}
\end{equation}
which we can read as the scalar product of two states which are known to be of finite norm, namely $\phi(f,t)\ket{0}$ (by assumption) and $\bra{\psi}\Phi(R_2)$ (which is the action of the well defined operator $\Phi(R_2)$ on a finite state). Then, tracing back our steps, we can compute the finite matrix elements of $\phi(f,t)$ between two arbitrary states. We deduce that $\phi(f,t)$ is a well defined operator on $\HH$, which concludes the argument.

\subsection{A possible solution}
\label{sec:solution}
In this section we summarize the situation and propose a possible solution of the difficulty.

First of all, fields (without time derivatives) admit a sharp time restriction in renormalizable theories treated in perturbation theory. This is because $\Gamma^{(2)}$ must have mass dimension 2, and since there are no couplings with negative mass dimension by assumption, $p^2$ can only appear either linearly or inside logarithms, which more generally indicate a dependence $\propto(p^2)^{-c\epsilon}$, for some number $c$, $\epsilon$ being the dim\nobreakdash-reg regulator. So ultimately the 2\nobreakdash-point function goes as $\frac{1}{(p^2)^{1+c\epsilon}}$ at large momentum, thus passing the requirement in \cref{eq2:existence_of_two_point_function}.

Non-perturbatively things are not so easy, and one should check the anomalous dimension of the field and \cref{eq2:existence_of_two_point_function} explicitly, because the mass dimension need not be close to $1$ anymore.
If we try to take the sharp time restriction of time derivatives of fields, we are always in trouble. All these results extend from the 2\nobreakdash-point function to any correlator, because of the argument given in \cref{sec:higher_point_functions_theorem}.
\smallskip

As we have seen in \cref{sec:time_smearing}, the most straightforward solution would be to smear our correlators in time (independently for each field), but the result would be quite cumbersome. So we rather ask: how do we make sense of E.T. correlators? From the general Wightman framework \cite{Streater:1989vi}, we know that these divergences are a distributional problem that arises in the coincident point limit (or more generally not at space-like separation), so one possibility would be to work in position instead of momentum space. The resulting correlator could then be trusted, as long as one never attempts to take any coincident points limit.
We only need the external legs of the correlator to be in position space, so all intermediate computations can still be performed (more conveniently) in momentum space. Actually we can take the momentum space E.T. correlator that we would normally compute at finite UV cutoff, perform a Fourier transform to position space for all the external legs and finally remove the UV\nobreakdash-regulator.

As an example, let us go back to the 2\nobreakdash-point function at finite $\Delta t,\,\Delta\vec x$
\begin{equation}
\label{eq:space_2pt}
    \braket{\phi(\vec x_1,t_1)\phi(\vec x_2,t_2)} = \frac{1}{2} \int\dd\mu^2\,\rho(\mu^2)\int\frac{\dd^3 p}{(2\pi)^3} \frac{1}{\Ep} e^{i\vec p\cdot\Delta \vec x-i\Ep \Delta t}
\end{equation}
Notice that the argument give right above \cref{eq2:existence_of_two_point_function} relied on the positivity of the integrand at E.T., which now is missing. Indeed at E.T. the space Fourier transform of $\frac{1}{2E_{\vec p}}$ is the free propagator at space-like separation, which enjoys a suppression proportional to $e^{-\mu |\Delta\vec x|}$ that makes the $\dd\mu^2\rho(\mu^2)$ integral always convergent at \emph{separate} points.

If we are interested in $\braket{\phi(\vec x_1,t_1)\dot\phi(\vec x_2,t_2)}$, we have instead
\begin{equation}
\label{eq:space_2pt_tder}
    \braket{\phi(\vec x_1,t_1)\dot\phi(\vec x_2,t_2)} = \frac{i}{2} \int\dd\mu^2\,\rho(\mu^2)\int\frac{\dd^3 p}{(2\pi)^3} e^{i\vec p\cdot\Delta \vec x-i\Ep \Delta t}
\end{equation}
thus at E.T. we get
\begin{equation}
\label{eq:space_2pt_tder}
    \braket{\phi(\vec x_1,t)\dot\phi(\vec x_2,t)} = \frac{i}{2} \delta^3(\vec x_1-\vec x_2)\int\dd\mu^2\,\rho(\mu^2)
\end{equation}
which is finite (actually zero) at distinct points.

\section{Examples}
\label{sec:examples}
This section is devoted to working out some flat space examples in detail. First, we will discuss the prototypical $\lambda\phi^4$ model, checking every statement made on general grounds in the previous section. We will try to elucidate in particular how the E.T. renormalized commutator $[\phi,\dot\phi]$ can be divergent, while the non-renormalized one remains finite. We continue in \cref{sec:composite_phi2} by considering a composite operator in a very simple (actually free) theory: this example is so simple that its de~Sitter counterpart can be easily studied as well. We conclude by touching on effective field theories, where the presence of an explicit cutoff calls an infinite time localization into question even more.

\subsection{Single scalar, $\lambda\phi^4$ model}
\label{sec:phi4_flat}
We now delve more into the details of the above arguments by studying an example. Our aim will be to show that in $d=4-\epsilon$, despite the divergence of correlation functions in the limit $\epsilon\rightarrow 0$, the commutator of the bare fields $[\phi_0,\dot\phi_0]$ is finite at equal times. Additionally, we will show that correlation functions of renormalized fields are divergent when evaluated at equal times. We consider $\lambda\phi^4$ theory
\begin{equation}
	\label{eq3:phi4_lagrangian}
	\LL = \frac{1}{2}(\partial_\mu\phi_0)^2-\frac{1}{2}m_0^2\phi_0^2-\frac{\lambda_0\mu^\epsilon}{4!}\phi_0^4
\end{equation}
in flat space. It is well known that, in dimensional regularization at two loops, the 2\nobreakdash-point function gets contributions from the following diagrams

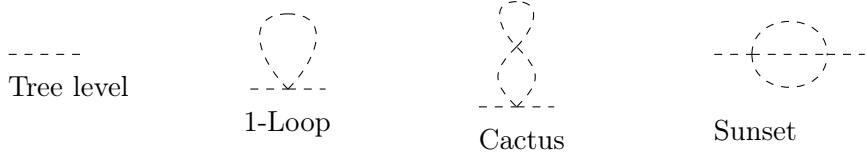
\begin{center}
%  tree level
\begin{minipage}{.2\textwidth}
\begin{tikzpicture}
\begin{feynman}
    \vertex (i1) at (0,0);
    \vertex (i2) at (1,0);
\diagram*{
    (i1) --  [scalar] (i2),
};
\end{feynman}
\end{tikzpicture}

Tree level
\end{minipage}
%  1 loop
\begin{minipage}{.2\textwidth}
\begin{tikzpicture}
\begin{feynman}
    \vertex (i1) at (0,0);
    \vertex (i2) at (1,0);
    \vertex (a) at (0.5,0);
    \vertex (b) at (0.5,1);
    \diagram*{
        (i1) -- [scalar] (a) -- [scalar] (i2),
        (a) -- [scalar, out=45, in=0, looseness=1.5] (b),
        (b) -- [scalar, out=180, in=135, looseness=1.5] (a),
    };
\end{feynman}
\end{tikzpicture}

1-Loop
\end{minipage}
%  cactus
\begin{minipage}{.2\textwidth}
\begin{tikzpicture}
\begin{feynman}
    \vertex (i1) at (0,0);
    \vertex (i2) at (1,0);
    \vertex (a) at (0.5,0);
    \vertex (b) at (0.5,0.8);
    \vertex (c) at (0.5,1.4);
    \diagram*{
        (i1) -- [scalar] (a) -- [scalar] (i2),
        (a) -- [scalar, out=45, in=315, looseness=1.5] (b),
        (b) -- [scalar, out=225, in=135, looseness=1.5] (a),
        (b) -- [scalar, out=45, in=0, looseness=1.5] (c),
        (c) -- [scalar, out=180, in=135, looseness=1.5] (b),
    };
\end{feynman}
\end{tikzpicture}

Cactus
\end{minipage}
%  sunset
\begin{minipage}{.2\textwidth}
\begin{tikzpicture}
\begin{feynman}
    \vertex (i1) at (0,0);
    \vertex (i2) at (2,0);
    \vertex (a) at (0.5,0);
    \vertex (b) at (1.5,0);
    \diagram*{
        (i1) -- [scalar] (a) -- [scalar] (b) -- [scalar] (i2),
        (a) -- [scalar, half right] (b),
        (a) -- [scalar, half left] (b),
    };
\end{feynman}
\end{tikzpicture}

Sunset
\end{minipage}
\end{center}

The resulting 2\nobreakdash-point amplitude is
\begin{align}
\begin{split}
\label{eq:phi4_gamma2}
    \Gamma^{(2)}_0 = p^2-m_0^2-\frac{\lambda_0 m_0^2}{2(4\pi)^2}\left(\frac{4\pi\mu^2}{m_0^2}\right)^{\frac{\epsilon}{2}}\Gamma\left(-1+\frac{\epsilon}{2}\right)+\\
    +\frac{\lambda_0^2m_0^2}{4(4\pi)^4}\left(\frac{4\pi\mu^2}{m_0^2}\right)^{\epsilon}\Gamma\left(-1+\frac{\epsilon}{2}\right)\Gamma\left(\frac{\epsilon}{2}\right)-\frac{\lambda_0^2\Gamma(\epsilon)}{6(1-\epsilon)(4\pi)^4}(4\pi\mu^2)^\epsilon(3m_0^2 A(p^2)+B(p^2)),\\
    A(p^2) = \int_0^1\dd x\int_0^1\dd y\, F(x,y,p^2),\quad
    B(p^2) = -p^2 \int_0^1\dd x\int_0^1\dd y\, y\, F(x,y,p^2)\\
    F(x,y,p^2) = [x(1-x)]^{-\frac{\epsilon}{2}}(1-y)y^{\frac{\epsilon}{2}-1}\left[-p^2y(1-y)+m_0^2\left(1-y+\frac{y}{x(1-x)}\right)\right]^{-\epsilon}
\end{split}
\end{align}
where the Feynman $i\varepsilon$ prescription can be implemented by giving a small negative imaginary part to $m_0^2$. The renormalization constants at order $\lambda^2$ in MS scheme are
\begin{align}
\begin{split}
	\phi_0 = Z^{1/2}\phi,\quad \lambda_0 = \frac{Z_\lambda}{Z^2}\lambda,\quad
    Z = 1-\frac{1}{12\epsilon}\frac{\lambda^2}{(4\pi)^4}\\
	Z_{m^2} = 1+\frac{\lambda}{(4\pi)^2\epsilon}+\frac{\lambda^2}{(4\pi)^4}\left(\frac{2}{\epsilon^2}-\frac{1}{2\epsilon}\right),\quad
	Z_\lambda = 1+\frac{3\lambda}{(4\pi)^2\epsilon}+\frac{\lambda^2}{(4\pi)^4}\left(\frac{9}{\epsilon^2}-\frac{3}{\epsilon}\right)
\end{split}
\end{align}

The 2\nobreakdash-point correlator is
\begin{align}
\label{eq:phi4_2pt_correlator}
\begin{split}
    \braket{\phi_0(E,\vec p)\phi_0(-E,-\vec p)}' = \frac{i}{\Gamma^{(2)}_0} = \frac{i}{p^2-m_0^2+i\varepsilon}+\\+\frac{i\lambda_0 m_0^2\,\Gamma\left(-1+\frac{\epsilon}{2}\right)}{2(4\pi)^2(p^2-m_0^2+i\varepsilon)^2}\left(\frac{4\pi\mu^2}{m_0^2}\right)^{\frac{\epsilon}{2}}
    +\frac{i\lambda_0^2 m_0^4\,\Gamma\left(-1+\frac{\epsilon}{2}\right)^2}{4(4\pi)^4(p^2-m_0^2+i\varepsilon)^3}\left(\frac{4\pi\mu^2}{m_0^2}\right)^{\epsilon}+\\
    -\frac{i\lambda_0^2m_0^2\,\Gamma\left(-1+\frac{\epsilon}{2}\right)\Gamma\left(\frac{\epsilon}{2}\right)}{4(4\pi)^4(p^2-m_0^2+i\varepsilon)^2}\left(\frac{4\pi\mu^2}{m_0^2}\right)^{\epsilon}
    +\frac{i\lambda_0^2\,\Gamma(\epsilon)\,(4\pi\mu^2)^\epsilon (3m_0^2 A(p^2)+B(p^2))}{6(1-\epsilon)(4\pi)^4(p^2-m_0^2+i\varepsilon)^2}
\end{split}
\end{align}
where the $i\varepsilon$ prescription was made explicit.

To make contact with the WFU results, we now go from energy to time domain
\begin{align}
    \braket{\phi_0(t,\vec p)\phi_0(0,-\vec p)}' = \int\frac{\dd E}{2\pi} e^{iE t} \braket{\phi_0(E,\vec p)\phi_0(-E,-\vec p)}'
\end{align}
For most of the terms in $\braket{\phi_0(E,\vec p)\phi_0(-E,-\vec p)}'$, the integration is straightforwardly performed. The tree level contribution is $\frac{e^{-iE_p|t|}}{2E_p}$, and it gives rise to the correct E.T. commutation relation for $[\phi,\dot\phi]$. All higher order contributions to $[\phi,\dot\phi]$ are expected to vanish. The 1-loop and cactus diagrams contribute with terms proportional to
\begin{equation}
    e^{-iE_p|t|}(1+iE_p|t|),\quad e^{-iE_p|t|}(3+3iE_p|t|-E_p^2t^2)
\end{equation}
where the coefficients contain $\frac{1}{\epsilon}$ poles. When taking $\left.\deriv{}{t}\right|_{0^+} - \left.\deriv{}{t}\right|_{0^-}$ to compute the contribution to the E.T. commutator, we get zero as expected. Finally, the sunset term containing the $A,\,B$ functions has branch cuts, making its evaluation non-trivial, and the cancellation of its contribution will be shown in \cref{app:sunset}.

The computation above suffices to show that $[\phi,\dot\phi]=i$ at E.T., but it is not very illuminating. Following the general strategy discussed in~\cref{sec:KL-general}, we now instead extract the spectral density $\rho_0(\mu^2)$ from $\braket{\phi_0(E,\vec p)\phi_0(-E,-\vec p)}'$ in \cref{eq:phi4_2pt_correlator}, bypassing the need to go to time domain. This is easily done because
\begin{equation}
    \lim_{\epsilon\rightarrow0^+}\frac{i}{p^2-m_0^2+i\varepsilon} = P\frac{i}{p^2-m_0^2}+\pi\delta(p^2-m_0^2)
\end{equation}
where $P$ indicates the principal part, so
\begin{equation}
    \Re[\braket{\phi_0(E,\vec p)\phi_0(-E,-\vec p)}'] = \int\dd\mu^2\rho_0(\mu^2) \Re\left[\frac{i}{p^2-\mu^2+i\varepsilon}\right] = \pi\rho_0(p^2)
\end{equation}
The tree level term gives the expected $\delta$ function centered on the particle mass, while all other terms except for the sunset contribution immediately give zero. For the sunset, besides picking up the residue at the pole $p^2=m_0^2$, we also get a term from the cut when the energy is above threshold ($p^2>(3m)^2$). Using
\begin{equation}
    (-x-i\varepsilon)^{-\epsilon} = x^{-\epsilon}(\cos(\pi\epsilon)+i\sin(\pi\epsilon)),\text{  if }x>0
\end{equation}
we get
\begin{align}
\begin{split}
    \rho_0(p^2) = \delta(p^2-m_0^2)+\frac{\lambda_0^2\,\Gamma(\epsilon)\,(4\pi\mu^2)^\epsilon (3m_0^2 A'(m_0^2)+B'(m_0^2))}{6(1-\epsilon)(4\pi)^4}\delta(p^2-m_0^2)+\\-
    \frac{\lambda_0^2\,\Gamma(\epsilon)\epsilon\,(4\pi\mu^2)^\epsilon}{6(1-\epsilon)(4\pi)^4(p^2-m_0^2)^2}\int_0^1\dd x\dd y[x(1-x)]^{-\epsilon/2}(1-y)y^{-1+\epsilon/2}\times\\
    \times(3m_0^2-p^2y)\left[p^2y(1-y)-m_0^2(1-y+\frac{y}{x(1-x)})\right]^{-\epsilon}\theta\left(p^2-\frac{m_0^2}{y(1-y)}(1-y+\frac{y}{x(1-x)})\right)
\end{split}
\end{align}
where $\theta$ is the Heaviside step function. The $p^2=m_0^2$ contribution has a $\frac{1}{\epsilon}$ pole, readily evaluated to be $-\frac{1}{12\epsilon}\frac{\lambda_0^2}{(4\pi)^4}\delta(p^2-m_0^2)$ as expected. The cut contribution does not have $\frac{1}{\epsilon}$ poles: to show this, notice the $y$ domain of integration is restricted to be always a finite distance away from 0 by the $\theta$ function, thus the problematic $y^{-1+\epsilon/2}$ term is actually harmless. For our purposes, it will be sufficient to evaluate the cut contribution in the limit of large $p^2$. Then $3m_0^2\ll p^2y$ and
\begin{equation}
    \rho_0(p^2) \simeq Z\delta(p^2-m_0^2) +\frac{\lambda_0^2}{12(4\pi)^4(p^2-m_0^2)^2} (p^2)^{1-\epsilon}
\end{equation}
where we dropped $\OO(\epsilon^0)$ terms in the coefficient of $\delta$, while the last term is approximated to leading order in large $p^2$. We can now check that
\begin{equation}
\label{eq:phi4_integral_rho0}
    \int_0^\infty\dd\mu^2\rho_0(\mu^2) = Z + \frac{\lambda_0^2}{12(4\pi)^4}\frac{1}{\epsilon} + \OO(\epsilon^0) = \OO(\epsilon^0)
\end{equation}
and the $\frac{1}{\epsilon}$ divergence cancelled, as announced in \cref{eq2:rho0_integral_1}. The pole and the cut contributions coming from the sunset actually exactly cancel, but showing this is beyond the scope of our computation. Notice the need to keep $\epsilon$ finite when it sits at the exponent of $p^2$: expanding it is a priori invalid when $p\rightarrow\infty$, and one can explicitly check that substituting $(p^2)^{-\epsilon}\rightarrow 1-\epsilon\ln p^2+\OO(\epsilon^2)$ would lead to insanable divergences in \cref{eq:phi4_integral_rho0}.
\smallskip

We now discuss the renormalized correlators. The mass is renormalized by $Z_{m^2}$, leading to $m_0^2\rightarrow m^2$ inside all formulas. Similarly, since $\lambda$ enters our formulas only as $\lambda^2$, we can set $Z_\lambda=1$ up to corrections that appear at higher loops. Finally $Z$ rescales $\phi_0$ and $\rho_0$. As claimed in \cref{sec:KL-general},
\begin{equation}
    \rho_R(\mu^2) = Z^{-1}\rho_0(\mu^2) \simeq \delta(\mu^2-m^2)+\frac{\lambda^2}{12(4\pi)^4(\mu^2-m^2)^2} (\mu^2)^{1-\epsilon}
\end{equation}
The integral $\int_0^\infty \rho_R(\mu^2)\dd\mu^2$ is now UV divergent, and the divergence is precisely the one predicted in~\cref{eq2:rho_integral_divergence}. On the other hand, the condition in \cref{eq2:existence_of_two_point_function} is satisfied even as $\epsilon\rightarrow0$, indicating that the E.T. $\braket{\phi\phi}$ correlator exists.

\subsection{Composite operator $:\phi^2(x):$ in a free scalar theory}
\label{sec:composite_phi2}
As pointed out in \cref{eq:existence_2pt_cft_condition}, the condition in \cref{eq2:existence_of_two_point_function} is not satisfied for CFT operators of dimension $\Delta>\frac{3}{2}$, indicating for example that $:\phi^2(x):$ in a free scalar theory should not admit a sharp time restriction because its mass dimension is $\Delta=2$. We verify this in this subsection. We suppress all time dependencies from operators and correlators, always working at fixed equal time.

We begin by observing that
\begin{equation}
    :\phi^2(\vec x): = \int \frac{\dd^3 p_1}{(2\pi)^3}\frac{\dd^3 p_2}{(2\pi)^3} e^{-i(\vec p_1 + \vec p_2)\cdot \vec x} :\tilde\phi(\vec p_1)\tilde\phi(\vec p_2): =
    \int \frac{\dd^3 p}{(2\pi)^3}\frac{\dd^3 P}{(2\pi)^3} e^{-i\vec P\cdot \vec x} :\tilde\phi(\vec p)\tilde\phi(\vec P-\vec p):
\end{equation}
where we separated the internal momentum and the centre of mass momentum. In momentum space, we can then define $:\tilde\phi^2(\vec P):$ as
\begin{equation}
    :\tilde\phi^2(\vec P): = \int \frac{\dd^3 p}{(2\pi)^3} :\tilde\phi(\vec p)\tilde\phi(\vec P-\vec p):
\end{equation}
Recall from \cref{eq2:flat_space_2correlator} applied to a free theory (or from the WFU approach) that
\begin{equation}
    \braket{\tilde\phi(\vec p_1)\tilde\phi^*(\vec p_2)}_t = \frac{1}{2E_{\vec p_1}}(2\pi)^3\delta^3(\vec p_1-\vec p_2)
\end{equation}
so
\begin{align}
\begin{split}
    \braket{:\tilde\phi^2(\vec P_1)::\tilde\phi^{2*}(\vec P_2):}_t = \int \frac{\dd^3 p_1}{(2\pi)^3} \frac{\dd^3 p_2}{(2\pi)^3} \braket{:\tilde\phi(\vec p_1)\tilde\phi(\vec P_1-\vec p_1)::\tilde\phi^*(\vec p_2)\tilde\phi^*(\vec P_2-\vec p_2):}_t =\\
    = 2\times (2\pi)^3 \delta^3(\vec P_1-\vec P_2) \int \frac{\dd^3 p_1}{(2\pi)^3}\frac{1}{2E_{\vec p_1}}\frac{1}{2E_{\vec P_1-\vec p_1}}
\end{split}
\end{align}
It is readily observed, from $E_{\vec p_1}\sim E_{\vec P_1-\vec p_1}\sim p_1$ as $p_1\rightarrow \infty$, that the last integral is linearly divergent. We stress that this happens despite normal ordering, which already subtracted all unphysical divergences and made $:\phi^2(x):$ a well-defined operator when suitably smeared against test functions: the problem lies entirely in forcing the correlator to be at equal times in 3-momentum space. As mentioned at the end of section~\ref{sec:interpretation}, the divergence is absent in position space for \emph{distinct} points at E.T., because clearly
\begin{equation}
    \braket{:\phi^2(\vec x)::\phi^2(\vec y):}_t = \frac{2}{[(\vec x-\vec y)^2]^{2}},\quad \vec x \neq \vec y
\end{equation}

This example with $:\phi^2(x):$ is simple enough that it can be easily studied even in de~Sitter. For simplicity, we restrict to the frequently studied case of a conformally coupled scalar field, with no interactions. Since we only used translational invariance, the whole discussion above still applies except that the 2\nobreakdash-point function for the elementary field $\phi$ now is
\begin{equation}
    \label{eq:dS_2pt_function}
    W_{\vec k}(\eta,\eta') = H^2\eta\eta'\frac{1}{2k}e^{ik(\eta-\eta')}
\end{equation}
so that
\begin{equation}
    \braket{:\tilde\phi^2(\vec P_1)::\tilde\phi^{2*}(\vec P_2):}_\eta
    = 2\times (2\pi)^3 \delta^3(\vec P_1-\vec P_2) \int \frac{\dd^3 p_1}{(2\pi)^3}W_{\vec p_1}(\eta,\eta)W_{\vec P_1-\vec p_1}(\eta,\eta)
\end{equation}
It is straightforward to see that the UV behavior of the integrand is the same, as one might have expected. Thus the same problem can also affect de~Sitter correlation functions.

\subsection{Light scalar, heavy scalar}
\label{sec:light-heavy_scalars}
Renormalizable theories such as the $\lambda\phi^4$ model discussed in \cref{sec:phi4_flat} are in principle valid up to arbitrarily large energies (aside for the Landau pole). On the contrary non-renormalizable theories stop making sense at energies above their cutoff and are only interpretable as effective field theories (EFTs), which prompts the question of what happens to the divergences in sharp time correlators in this context.

The first general point that we want to raise is the failure of the argument given in section~\ref{sec:solution}, according to which $\rho(\mu^2)$ scales as $\sim(p^2)^{1-c\,\epsilon}$ at large momentum. Indeed, it is known that $\rho(\mu^2)$ does not exhibit sub-exponential growth in non-renormalizable theories, making virtually every equation in section~\ref{sec:KL-general} ill-defined (we will see in \cref{sec:light-heavy_scalars_leading-terms} why, with the appropriate interpretation, this is not a problem for EFTs).

The failure of formulas derived from an axiomatic framework comes as no surprise given the finite regime of validity of the theory, and the reader should rightfully ask what happens in practice when doing computations within the usual EFT framework. We will now work through an explicit example to show that higher derivative terms, generic in an EFT Lagrangian, spoil the existence of sharp time restrictions of fields even in perturbation theory, making correlation functions divergent in 3\nobreakdash-momentum space. Next, we will show that a resummation of diagrams can give finite results in specific cases, which however are hard to interpret physically because of the appearence of spurious contributions. Indeed, while in renormalizable theories only coincident points are problematic at equal times, we will see in \cref{eq:EFT_cutoff} that in theories with a cutoff $\Lambda$ any time localization more precise than $\frac{1}{\Lambda}$ is problematic.
\medskip

Consider a model consisting of a light scalar $\pi$ and a heavy scalar $\phi$ \footnote{This model was chosen to avoid the appearence of logarithms in \cref{eq:EFT_Gamma2_schematic}, which makes the subsequent discussion simpler while illustrating the main point.}, coupled together
\begin{equation}
    \LL_{UV} = \frac{1}{2}(\partial_\mu \pi)^2 - \frac{1}{2}m^2 \pi^2 + \frac{1}{2}(\partial_\mu \phi)^2 - \frac{1}{2}M^2 \phi^2 +\frac{g}{2}\pi^2\phi
\end{equation}
At energies much below $M$, $\pi$ is described by the effective lagrangian
\begin{equation}
    \LL_{EFT} = \frac{1}{2}(\partial_\mu \pi)^2 - \frac{1}{2}m^2 \pi^2 + \dots
\end{equation}
where the dots encapsulate all higher order operators, which can be fixed by imposing matching conditions between $\LL_{UV}$ and $\LL_{EFT}$. In particular, higher derivative operators like (schematically) $\frac{g^2}{M^4}\partial^2\pi\partial^2\pi$ appear in $\LL_{EFT}$.

We will now show from a top-down computation that such terms proportional to $p^4$ dominate the high-energy limit of the 2\nobreakdash-point function $\Gamma^{(2)}_{EFT}$, computed to some finite order in $g$ and $\frac{1}{M}$. These terms will be important in what follows.
\medskip

In the UV theory, the 2\nobreakdash-point function of $\pi$ at 1 loop (in $\overline{MS}$ after renormalization) is
\begin{align}
\begin{split}
    i\Gamma^{(2)}_\pi = i(p^2-m^2)+(ig)^2\int\frac{\dd^4k}{(2\pi)^4} \frac{i}{k^2-m^2}\frac{i}{(p+k)^2-M^2}+\text{counter-terms} \simeq \\
    = i(p^2-m^2) +i\frac{g^2}{16\pi^2}\left[1+\ln\frac{\mu^2}{M^2}+\frac{p^2}{2M^2}+\frac{m^2}{M^2}\ln\frac{m^2}{M^2}+\right.\\
    \left.+\frac{p^4}{6M^4}+\frac{3m^2p^2}{2M^4}+\left(\frac{m^4}{M^4}+\frac{m^2p^2}{M^4}\right)\ln\frac{m^2}{M^2}+\OO\left(\frac{1}{M^6}\right)\right]+\OO(g^4)
\end{split}
\label{eq:EFT_gamma2}
\end{align}

The term $\frac{g^2}{16\pi^2}\frac{p^4}{6M^4}$ dominates the 2\nobreakdash-point function if we (illegitimately) extrapolate the theory up to large momenta.

We are now presented with a choice. We can either discuss the 2\nobreakdash-point correlator to order $g^2$, or we can resum the geometric series of $\OO(g^2)$ 1\nobreakdash-PI diagrams and study $\frac{i}{\Gamma^{(2)}}$.

\subsubsection{Leading terms}
\label{sec:light-heavy_scalars_leading-terms}
In the first case the 2\nobreakdash-point correlator is
\begin{equation}
\label{eq:EFT_correlator_leading-terms}
    \braket{\phi(\vec p)\phi(-\vec p)}'_t = i\int\frac{\dd E}{2\pi} \left(\frac{1}{p^2-m^2}+g^2\frac{-\frac{1}{16\pi^2}\frac{p^4}{6M^4}+\OO(p^2,p^0)}{(p^2-m^2)^2}\right)
\end{equation}
Due to the $p^4$ term, which dominates at large energies, the second part of the integrand gives a divergent contribution. This is to be contrasted with the $\lambda\phi^4$ example presented in \cref{sec:phi4_flat}, where the E.T. $\braket{\phi\phi}$ correlator existed. More generally, the presence of a $p^4$ term is only possible in a non-renormalizable theory, on dimensional grounds.

If we were to introduce a time smearing that cuts off frequencies higher than $\Omega$, the correlator \cref{eq:EFT_correlator_leading-terms} would get a contribution proportional to $g^2\frac{\Omega}{M^4}$, thus linearly divergent when the smearing is removed. This divergence signals a sensitivity of the result on the effective UV cutoff $\Omega$ introduced by the smearing, which is hardly acceptable in an EFT.

From this example it is also clear that $\rho(\mu^2)$ truncated to \emph{finite} order in the EFT couplings will still exhibit sub-exponential growth in any EFT, thus sidestepping some negative remarks presented at the beginning of \cref{sec:light-heavy_scalars}.

\subsubsection{Resummed series}
\label{sec:light-heavy_scalars_resummed}
The above treatment might seem a bit naive, because usually we think of inverting $\Gamma^{(2)}$ to determine the 2\nobreakdash-point correlator, which corresponds to resumming the geometric series of 1PI 2\nobreakdash-point diagrams. We should then discuss
\begin{equation}
\label{eq:sharp_time_from_Gamma}
    \braket{\phi\phi}_t = i\int\frac{\dd E}{2\pi} \frac{1}{\Gamma^{(2)}}
\end{equation}
directly. To simplify the notation, let us define $\tilde m,c_1,c_2$ such that $\Gamma^{(2)}$ in \cref{eq:EFT_gamma2} can be expressed as
\begin{equation}
\label{eq:EFT_Gamma2_schematic}
    \Gamma^{(2)} = p^2-\tilde m^2 +c_1 p^2+c_2 p^4,\quad c_1,c_2>0
\end{equation}
The integrand now has 4 poles, and in particular we can write
\begin{align}
\begin{split}
    \frac{1}{\Gamma^{(2)}} = \frac{1}{c_2(a-b)}\left(\frac{1}{p^2-a}-\frac{1}{p^2-b}\right)\\
    a = \frac{\sqrt{(c_1+1)^2+4 c_2 m^2}-c_1-1}{2 c_2},
    \quad b = -\frac{\sqrt{(c_1+1)^2+4 c_2 m^2}+c_1+1}{2 c_2}\\
    a \simeq m^2,\quad b\simeq -\frac{1}{c_2},\quad \frac{1}{c_2(a-b)}\simeq 1\text{ as }g\rightarrow0
\end{split}
\end{align}
which makes manifest the presence of an unphysical degree of freedom.

Reintroducing $i\varepsilon$, the integral in \cref{eq:sharp_time_from_Gamma} can now be trivially performed resulting in $\propto\frac{1}{E_a}-\frac{1}{E_b}$, which is the sum of simple free propagators. In particular, the mass squared $b$ is negative, leading to
\begin{equation}
    \frac{1}{E_b} = \frac{1}{\sqrt{\vec p^2+b}}\simeq i c_2\propto \frac{g^2}{M^4}
\end{equation}
which is a spurious contribution of the same order as the terms we kept in the EFT lagrangian. Going back once more to the idea of introducing a time smearing that cuts off frequencies above $\Omega$, we see that a reasonable consistency requirement we should impose is that the unphysical mass squared $b$ should be much larger than the smearing cutoff $\Omega^2$, leading to
\begin{equation}
\label{eq:EFT_cutoff}
    \Omega^2\ll \frac{M^4}{g^2}
\end{equation}
which is the simple statement that the energies we consider should be well below the strong coupling scale. The conclusion is that any time localization sharper than the theory cutoff, implicit in a WFU computation, is illegitimate in an EFT.

\section{Conclusions}
In this paper we examined the existence of sharp time restrictions for quantum fields, required to compute correlators of fields at equal time. This seemingly abstract and harmless assumption, at the basis of any Schroedinger-like picture of QFT, is actually known to be problematic. This is to be contrasted with the usual correlators expressed in terms of \emph{energy} (instead of time) and momentum, where this problem does not exist. This could call into question the validity of computations performed using the wavefunction of the universe (WFU) approach. We were therefore prompted to determine under which conditions the sharp time restriction is valid.

Fortunately, for elementary scalar fields, the condition we found in \cref{eq2:existence_of_two_point_function}, which in \cref{eq:existence_2pt_cft_condition} was reduced to the check that the UV mass dimension of the field should be at most $\frac{3}{2}$, is verified for free theories as well as scalar theories treated in perturbation theory (because the mass dimension is arbitrarily close to $1$). This check, while only guaranteeing the existence of the 2\nobreakdash-point function, extends to all $n$\nobreakdash-point correlators thanks to the argument given in \cref{sec:higher_point_functions_theorem}. The WFU results are then completely fine to all loop orders for elementary scalar fields.

However it is not hard to construct simple examples where equal time correlation functions are ill-defined: we showed how divergences can arise both in the presence of time derivatives and composite operators, even for very well known theories like $\lambda\phi^4$. Such divergences typically only appear at two loops, which is more than what generally considered for cosmological correlators. Despite the difficulties in performing explicit computations, there are some all-loop orders results like the flat space analogues (we will later comment about cosmology) of the cosmological optical theorem and cutting rules in WFU \cite{Meltzer:2021zin,Goodhew:2020hob,Goodhew:2021oqg,Baumann:2021fxj,Jazayeri:2021fvk,Melville:2021lst,Lee:2023kno}. These are, in our view, best understood as statements about the UV\nobreakdash-regulated theory, because the relevant correlators could in principle involve such time derivatives or composite operators.

While we only focused on scalar fields for simplicity, extensions to spinning fields are straightforward. For spin $1/2$ fields for example, an analogue of the divergences we found for $:\phi^2(x):$ could be the non-canonical terms that Schwinger discovered \cite{Schwinger:1959xd} in current commutators, which make sharp time restrictions of the electric current ill-defined already in a free theory (see \cite{Strocchi:2013awa} for a review). This example is instructive because it shows that even current commutators might not be finite at E.T., despite the role of conserved currents as symmetry generators (see~\cite{Strocchi:2013awa} for a solution to this problem).

The divergences we discussed only arise in the coincident space\nobreakdash-point limit of the equal time correlators, so a strategy to obtain a finite result could be to work in position rather than 3\nobreakdash-momentum space, at the price of obscuring the overall 3\nobreakdash-momentum conservation. For effective field theories this strategy, while formally working, is not free of issues because the infinite time localization requires that we know the correlation function at energies well above the EFT cutoff.

Finally, we draw some conclusions regarding cosmology. The first point is that our arguments could only apply to correlators at finite (and equal) conformal time $\eta$. This is because the more physically interesting late time limit $\eta\rightarrow0$ could also be achieved by computing the correlator at different $\eta_i$ for each operator $\OO_i$, and then sending each $\eta_i$ to zero independently. It is then unclear whether removing the UV cutoff after taking $\eta\rightarrow0$ in an equal time correlator would give the correct result. However we notice how this late time limit cannot be directly taken in the presence of IR divergences: in this case the UV divergences that one wishes to renormalize away would get mixed with the more artificial UV divergences we focused on (and with the IR divergences themselves, depending on the regulator), complicating the problem.

Secondly, although much of our discussion was carried out in flat space, we expect many of our results to carry over to de~Sitter since we were always concerned with UV properties of the correlators, which should be robust under IR deformations of the spacetime and operators like the Riemann tensor getting a vacuum expectation value. We also highlight that many of our arguments employed tools, like the \KL representation or the Reeh-Schlieder theorem, which have been successfully carried over to de~Sitter \cite{Bros:1994dn,Cacciatori:2024zrv,Loparco:2023rug,Bros:1995js}, further closing the gap between this work and the more interesting cosmological scenario. Similarly, while in cosmology we are interested in in\nobreakdash-in rather than in-out correlators, we do not expect significant deviations because in flat space the two correlators coincide (as shown in \cref{app:WFU_In-In_In-Out}).

\section{Aknowledgements}
The author would like to thank Manuel Loparco, Ugo Moschella, Enrico Pajer, Guilherme Pimentel and the participants of the Cosmological Correlators Workshop in Cortona for useful discussions. The author also expresses his gratitude to Franco Strocchi for useful discussions, especially regarding~\cref{sec:higher_point_functions_theorem}, and for his courses on quantum field theory.

\appendix
\section{Relation between WFU, In-In and In-Out}
\label{app:WFU_In-In_In-Out}
The purpose of this paper was to study in\nobreakdash-in correlators, but we only ever computed in-out correlators. The goal of this appendix is to recall why this is identical in flat space. We do this by first showing the general equivalence between the in\nobreakdash-in (also called Schwinger-Keldish) formalism and the wavefunction of the universe approach. Then we show that in flat space in\nobreakdash-in and in-out coincide.
\medskip

Recall the definition of an in\nobreakdash-in correlator of some string of fields $\Phi(t)$ on the state $\ket{\Omega_t}$, which is the time evolution of the vacuum $\ket{0}$ in the asymptotic past.
\begin{equation}
    \braket{\Phi(t)}_{\Omega_t} = \braket{0|\bar T e^{i\int_{-\infty}^t \dd t' H_{int}(t')}\Phi(t)T e^{i\int_{-\infty}^t \dd t' H_{int}(t')}|0}
\end{equation}
where $\bar T$ stands for anti time ordering. We can rewrite this as
\begin{equation}
    \braket{\Phi(t)}_{\Omega_t} = \braket{\Omega_t|\Phi(t)|\Omega_t} = \int \mathcal{D}\phi \braket{\Omega|\phi}\Phi(t)\braket{\phi|\Omega}
\end{equation}
where we used the identity $\bone = \int\mathcal{D}\phi \ket{\phi}\bra{\phi}$ and that $\ket\phi$ are eigenstates of the field operators, so that $\Phi(t)$ inside the integral is a number. Recalling the definition of wave function(al) $\Psi_\Omega(\phi) = \braket{\phi|\Omega}$, we finally get
\begin{equation}
    \braket{\Phi(t)}_{\Omega_t} = \int\mathcal{D}\phi |\Psi_\Omega(\phi)|^2 \Phi(t)
\end{equation}
which is the usual Born rule adopted in the WFU approach, proving the equivalence between this method and Schwinger-Keldish.
\medskip

We now come to the equivalence of in\nobreakdash-in and in-out formalisms in flat space, repeating a well known argument presented for example in \cite{Kamenev_2023,Donath:2024utn}.
We start by writing the in\nobreakdash-in correlator as
\begin{equation}
    \braket{\Phi(t)}_{\Omega_t} = \braket{0|U_{-\infty,t}\Phi(t) U_{t,-\infty}|0}
\end{equation}
Assuming now that we are in flat space, it is a well known fact \cite{Coleman:2011xi} that
\begin{equation}
\label{eq:adiabatic_vacuum}
    S\ket{0}=U_{+\infty,-\infty}\ket{0} = e^{i\vartheta}\ket{0}
\end{equation}
where the UV divergent phase is the sum of disconnected diagrams. Then we deduce
\begin{equation}
    \braket{\Phi(t)}_{\Omega_t} = \frac{\braket{0|U_{+\infty,-\infty}U_{-\infty,t}\Phi(t) U_{t,-\infty}|0}}{e^{i\vartheta}} = \frac{\braket{0|U_{+\infty,t}\Phi(t)U_{t,-\infty}|0}}{\braket{0|S|0}}
\end{equation}
which is precisely the usual in-out correlator formula. The crucial step of the proof is \cref{eq:adiabatic_vacuum}, which does not hold in a spacetime without time translational symmetry such as de~Sitter.

\section{Sunset diagram}
\label{app:sunset}
The contribution of the sunset diagram to $\braket{\phi\phi}'$ in $\lambda\phi^4$ theory is
\begin{equation}
\label{appsunset:correlator}
    S(p^2)=\frac{i\lambda_0^2\,\Gamma(\epsilon)\,(4\pi\mu^2)^\epsilon (3m_0^2 A(p^2)+B(p^2))}{6(1-\epsilon)(4\pi)^4(p^2-m_0^2+i\varepsilon)^2}
\end{equation}
and the goal of this appendix will be to Fourier transform this quantity to time domain
\begin{equation}
\label{appsunset:fourier}
    \int\frac{\dd E}{2\pi}S(p^2) e^{i E t}
\end{equation}

\begin{figure}
    \centering
    \begin{tikzpicture}
    \draw[blue, thick] (-1,0) -- (3,0);
    \path [draw=gray,snake it] (-1,0.3) -- (0.5,0.3);
    \path[draw=green,->] (-1,0.1) -- (0.5,0.1);
    \draw[draw=green] (0.5,0.1) arc (-90:90:0.195);
    \path[draw=green,->] (0.5,0.5) -- (-1,0.5);
    \draw[red] (1.3,0.4) -- (1.5,0.6);
    \draw[red] (1.5,0.4) -- (1.3,0.6);
    \draw[draw=green,->] (1.4,0.2) arc (-90:245:0.3);
    \draw[draw=green,->] (-1.1,0.5) arc (167:0:2);
    \end{tikzpicture}
    \caption{Evaluation of the sunset diagram in $\lambda\phi^4$ theory. We display the upper complex $E$ plane. The real line, in blue, is the integration domain. When $t>0$ in \cref{appsunset:fourier}, the integration domain can be deformed to the green lines.}
    \label{fig:sunset_cut}
\end{figure}
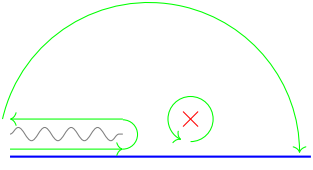
What is the analytic structure of $S(p^2)$ in the complex $E$ plane? There are two poles at $E=\pm (E_p-i\varepsilon)$, and from the definitions of $A(p^2)$, $B(p^2)$ in \cref{eq:phi4_gamma2} we see two branch cuts emanating from values of $E$ given by (see the definition of $F(x,y,p^2)$ in \cref{eq:phi4_gamma2})
\begin{equation}
    -(E^2-\vec p^2)y(1-y)+m_0^2\left(1-y+\frac{y}{x(1-x)}\right)-i\varepsilon=0
\end{equation}
These branch points are symmetric with respect to the origin of the complex $E$ plane, and the one with negative real part has a small positive imaginary part.

When $t>0$, we can close the contour of \cref{appsunset:fourier} in the upper half plane, picking up the energy pole. To fully close the contour however, we have to go around the cut and account for its contribution. The end result is described in \cref{fig:sunset_cut}, so we have to evaluate the cut and the pole contributions separately.

The pole contributes, up to $\OO(\epsilon^0)$ terms, as
\begin{equation}
    -\frac{\lambda_0^2\, (3m_0^2 A(m_0^2)+B(m_0^2))}{6\epsilon(4\pi)^4}\frac{1+iE_p|t|}{4E_p^3} e^{-iE_p|t|} + \frac{\lambda_0^2\, (3m_0^2 A'(m_0^2)+B'(m_0^2))}{6\epsilon(4\pi)^4} \frac{e^{-iE_p|t|}}{2E_p}
\end{equation}
The contribution to E.T. $\braket{\phi\dot\phi}'$ then comes only from $B'(m_0^2)$, to leading order in $\frac{1}{\epsilon}$, and it is
\begin{equation}
    -\frac{i}{4} \frac{\lambda_0^2}{6\epsilon(4\pi)^4}
\end{equation}
The cut can be handled recalling
\begin{equation}
    (-x-i\varepsilon)^{-\epsilon} = x^{-\epsilon}(1+i\epsilon\pi)
\end{equation}
so that the discontinuity along the cut is $x^{-\epsilon}2\pi i \epsilon$. Then the cut contributes as
\begin{align}
\begin{split}
    \frac{i\lambda_0^2}{6\epsilon(4\pi)^4}\int_{-\infty}^0\frac{\dd E}{2\pi}\frac{e^{iEt}}{(p^2-m_0^2+i\varepsilon)^2}\int_0^1\dd x \dd y\,
    (3m_0^2-p^2y)[x(1-x)]^{-\epsilon/2}(1-y)y^{-1+\epsilon/2}\times\\
    \times\left[p^2y(1-y)-m_0^2\left(1-y+\frac{y}{x(1-x)}\right)\right]^{-\epsilon}2\pi i \epsilon\,\theta\left[p^2y(1-y)-m_0^2\left(1-y+\frac{y}{x(1-x)}\right)\right]
\end{split}
\end{align}
where the integration domain is first restricted to $(-\infty,0)$ to discard the other cut, and then the $\theta$ function selects the correct half-line, whose origin is $x,y$ dependent.

The reader should notice that the pole from $\Gamma(\epsilon)$ cancels against the $2\pi i \epsilon$ coming from the discontinuity; moreover the $x,y$ integral cannot give a divergence as $\epsilon\rightarrow0$, because the dangerous point $y=0$ is never part of the integration domain for any finite $p^2$.

Thus we can extract the leading $\frac{1}{\epsilon}$ behavior of the integral by taking $\epsilon\rightarrow0$ inside the integrand, except for the factor which depends on $p^2$. Focusing on the leading behavior, we get
\begin{equation}
    \sim -\frac{\lambda_0^2}{6(4\pi)^4}\int_{-\infty}\dd E\frac{e^{iEt}}{(p^2-m_0^2+i\varepsilon)^2}\frac{(-p^2)}{2}
    \left[p^2\right]^{-\epsilon}
\end{equation}
while the contribution to E.T. $\braket{\phi\dot\phi}'$ is
\begin{equation}
    \sim -i\frac{\lambda_0^2}{12(4\pi)^4}\int_{-\infty}\dd E\,E\frac{1}{(p^2-m_0^2+i\varepsilon)^2}p^2 \left[p^2\right]^{-\epsilon}
    \sim i\frac{\lambda_0^2}{24(4\pi)^4\epsilon}
\end{equation}

To evaluate the correlator for $t<0$ we would close the contour in the lower half $E$ plane, obtaining completely analogous results. Indeed the E.T. $\braket{\dot\phi\phi}$ cut and pole contributions are opposite to the ones obtained here for $\braket{\phi\dot\phi}$, so we simply double them when considering the commutator. This implies that for bare fields $[\phi_0,\dot\phi_0]=\OO(\epsilon^0)$, so the $\frac{1}{\epsilon}$ pole cancels, while at E.T.
\begin{equation}
    \braket{\phi_R\dot\phi_R}'=-\braket{\dot\phi_R\phi_R}'=\frac{1}{2}\braket{[\phi_R,\dot\phi_R]}' = i\frac{\lambda^2}{24(4\pi)^4\epsilon}+\OO(\epsilon^0)
\end{equation}
as expected.

\printbibliography[heading=bibintoc]

\end{document}